\documentclass[twocolumn,showpacs, superscriptaddress, groupedaddress,nofootinbib]{revtex4}  
\usepackage{graphicx}  
\usepackage{dcolumn}   
\usepackage{bm}        
\usepackage{amssymb, amsmath}
\usepackage{wasysym}
\usepackage{soul}

\def\beq{\begin{equation}}
\def\eeq{\end{equation}}
\usepackage[usenames,dvipsnames,svgnames]{xcolor}
\usepackage{hyperref}
\hypersetup{colorlinks, citecolor=black, linkcolor=black, urlcolor=black}

\usepackage{amsmath}
\usepackage{amsfonts}
\usepackage{amssymb}
\usepackage{bbm}                
\usepackage{graphicx, rotating}
\usepackage{epstopdf}
\usepackage{epsfig}
\usepackage{latexsym}
\usepackage{graphicx}
\usepackage{bbold}
\usepackage{color}
\usepackage{amsmath,amssymb}

\begin{document}
\widetext
\begin{flushright}
	IPPP/17/79
\end{flushright}
\title{The Nelson-Barr Relaxion}
\author{Oz Davidi}
\affiliation{Department of Particle Physics and Astrophysics, Weizmann Institute of Science, Rehovot 7610001,Israel}
\author{Rick S. Gupta}
\affiliation{Department of Particle Physics and Astrophysics, Weizmann Institute of Science, Rehovot 7610001,Israel}
\affiliation{Institute for Particle Physics Phenomenology, Department of Physics, Durham University, DH1 3LE, Durham, United Kingdom}
\author{Gilad Perez}
\affiliation{Department of Particle Physics and Astrophysics, Weizmann Institute of Science, Rehovot 7610001,Israel}
\author{Diego Redigolo}
\affiliation{Department of Particle Physics and Astrophysics, Weizmann Institute of Science, Rehovot 7610001,Israel}
\affiliation{Raymond and Beverly Sackler School of Physics and Astronomy, Tel-Aviv University, Tel-Aviv 69978, Israel}
\author{Aviv Shalit}
\affiliation{Department of Particle Physics and Astrophysics, Weizmann Institute of Science, Rehovot 7610001,Israel}
\begin{abstract}
Cosmological relaxation models in which the relaxion is identified with the QCD axion, generically fail to account for the smallness of the strong CP phase. We present a simple alternative solution to this ``relaxion CP problem''  based on the Nelson-Barr mechanism. We take CP to be a symmetry of the UV theory, and the relaxion to have no anomalous coupling with QCD. The non-zero vacuum expectation value of the relaxion breaks CP spontaneously, and the resulting phase is mapped to the Cabibbo-Kobayashi-Maskawa phase of the Standard Model. The extended Nelson-Barr quark sector generates the relaxion ``rolling'' potential radiatively, relating the new physics scale with the relaxion decay constant. With no new states within the reach of the LHC, our relaxion can still be probed in a variety of astrophysical and cosmological processes, as well as in flavor experiments. 
\end{abstract}
\maketitle

The large hierarchy between the electroweak~(EW) scale and the Planck scale, and the smallness of the strong CP phase compared to the Cabibbo-Kobayashi-Maskawa~(CKM) phase are two of the main mysteries of modern particle physics. In this paper, we show that the two problems can be explained in a unified way by combining the framework of cosmological relaxation~\cite{Graham:2015cka} with the Nelson-Barr~(NB) mechanism~\cite{Nelson:1983zb,Barr:1984qx,Barr:1984fh}.  

The key feature of cosmological relaxation is to make the Higgs mass dependent on the cosmological evolution of the relaxion. The relaxion is the pseudo Nambu-Goldstone boson of a spontaneously broken abelian symmetry $U(1)_{\text{clock}}$ that gets explicitly broken by two sequestered sectors with exponentially hierarchical charges. These sectors feed into the relaxion potential, generating exponentially different periodicities. After the relaxion  dynamics is resolved, the ratio of the two periodicities is related to the hierarchy between the EW scale and the UV cutoff, where extra new physics stabilizing the Higgs mass is expected. 

The structure of the relaxion potential forces it to get a vacuum expectation value~(VEV), introducing a CP violating phase in the theory~\cite{Graham:2015cka,Flacke:2016szy}. This property causes the ``relaxion CP problem'' - the difficulty of identifying the relaxion with the QCD axion, and connecting the strong CP problem with the naturalness of the electroweak scale. In this paper, we show that the relaxion CP problem turns into a blessing if the large periodicity of the relaxion potential is generated via a NB model, such as the ones described in Refs.~\cite{Bento:1991ez,Dine:2015jga}.

Following the NB basic setup, we assume that the UV theory preserves CP, and that the $U(1)_{\text{clock}}$ has zero Adler-Bell-Jackiw~(ABJ) anomaly with QCD. A discrete symmetry ensures that the CP violating phase, controlled by the relaxion VEV, is mapped into the CKM phase, while the strong CP phase remains zero at tree-level. In order to keep the strong CP phase below the bounds from measurements of the neutron dipole moment~\cite{Afach:2015sja}, the NB sector should be feebly coupled to the source of CP violation, so that radiative corrections are under control~\cite{Dine:2015jga}.

The NB relaxion solves the strong CP problem together with the hierarchy problem by means of the dynamics of a single light degree of freedom. This should be contrasted with more conventional solutions of the hierarchy problem, like Supersymmetry~(SUSY) or Compositeness, where a tension with CP observables is often present (see for example Refs.~\cite{Polchinski:1983zd,Buchmuller:1982ye, Dugan:1984qf,Agashe:2004ay}), and the strong CP problem is typically addressed by adding new degrees of freedom disconnected from the Higgs sector. 
 
Once the NB sector is specified, the relaxion potential with large periodicity is computable, and the UV cutoff gets connected to the relaxion decay constant. This provides a unique interplay between the LHC phenomenology, bounding the UV cutoff from below, and the low energy phenomenology of the relaxion itself. We show how the LHC bounds together with a successful solution of the strong CP problem imply that the spontaneous breaking of $U(1)_{\text{clock}}$ should happen at  a high scale. As a consequence, the NB relaxion phenomenology is generically pretty elusive, even though a non negligible relaxion-Higgs mixing can be probed in astrophysical and cosmological processes, and in flavor factories~\cite{Flacke:2016szy}.  

At the end of this letter, we discuss how the spontaneous breaking of $U(1)_{\text{clock}}$ at a high scale poses a ``relaxion quality problem'', which is  related to the theoretical challenge of screening Planck-suppressed effects in order to preserve the peculiar structure of the relaxion potential. 
\medskip

\section{The Relaxion CP Problem}
We start by  reviewing the main ingredients of the relaxion construction. The relaxion potential gets contributions from two  different sectors: the ``rolling'' sector and the ``backreaction'' sector, which we parametrize as 
\begin{align} 
V_{\text{roll}}&\!=\!\mu^2(\phi) H^\dagger\!H+\lambda_{H} (H^\dagger\!H)^2-r^2_{\text{roll}}\Lambda^{4}_H\cos\frac{\phi}{F} \,,\label{mu1}\\
\mu^2(\phi)&=\kappa \Lambda^2_H-\Lambda^2_H \cos \frac{\phi}{F} \,, 
\label{mu2}
\end{align}
and 
\begin{equation}
V_{\text{br}}=-M_{\text{br}}^{2}H^\dagger\!H \cos \frac{\phi}{f}-r^2_{\text{br}} M_{\text{br}}^{4} \cos \frac{\phi}{f}\,,\label{eq:back}
\end{equation}
where $H$ is the Standard Model (SM) Higgs doublet, and $\phi$ is the relaxion. $F$ is the ``large'' periodicity of the rolling sector, $f$ is the ``small'' periodicity of the backreaction sector, $\Lambda_H$ is the UV cutoff of the Higgs effective field theory, and $M_{\text{br}}$ parametrizes the mass threshold controlling the backreaction potential. We also introduced the coefficient $\kappa\lesssim1$ in front of the UV threshold to the Higgs mass, and the coefficients $r_{\text{roll}}$ and $r_{\text{br}}$ which account for the possible differences between the Higgs-dependent contributions and  the Higgs-independent contributions in the relaxion potential. Both coefficients satisfy $r_{\text{roll}}\,, r_{\text{br}}\gtrsim 1/4\pi\,$. The lower bound on $r_{\text{roll}}$ can easily be seen by closing the Higgs loop in the Higgs-dependent contributions in Eq.~\eqref{mu1}, while we refer to Refs.~\cite{Flacke:2016szy,Gupta:2015uea,Espinosa:2015eda} for a discussion on how an analogous bound is obtained for the backreaction sector.

During inflation, the relaxion rolls down the potential from some initial field value $\phi \lesssim \phi_c\approx -|F \cos^{-1} \kappa|\,$, such that $\mu^2(\phi)>0\,$. While rolling down, the relaxion dissipates energy through Hubble friction. Once $\phi\gtrsim \phi_c\,$, $\mu^2(\phi)<0$ and the SM Higgs doublet gets a VEV, breaking the EW symmetry. 

In the EW broken phase, the backreaction potential $V_{\text{br}}$ should generate  wiggles allowing the relaxion to stop where the Higgs mass is at its measured value (see Ref.~\cite{Graham:2015cka} for further details). This requirement gives an upper bound on the amplitude of the Higgs-independent contribution to the backreaction potential, which can be translated into an upper bound on the backreaction scale
\begin{equation} 
M_{\text{br}}\lesssim \frac{v}{r_{\text{br}}}\lesssim 4\pi v\,.\label{eq:boundback}
\end{equation}
If the above condition is fulfilled, the relaxion stops at $\phi_0=\langle \phi\rangle\,$, where $\vert\mu(\phi_0)\vert=m_h\,$, and~\cite{Gupta:2015uea}
\begin{align}
&\quad \sin \frac{\phi_0}{f}\sim\sin \frac{\phi_0}{F}\sim\mathcal{O}(1)\,,\label{eq:order1}\\
&\quad  \partial_{\phi} V=0\quad\Rightarrow\quad \frac{\Lambda_H}{ \Lambda_{\text{br}}} \sim \left(\frac{F}{r^2_{\text{roll}} f}\right)^{1/4}\,,\label{eq:deriv}
\end{align}
where $\Lambda_{\text{br}}=\sqrt{M_{\text{br}}v}$ is  the backreaction scale. 

Eq.~\eqref{eq:deriv} shows that a large ratio between the Higgs bare mass $\Lambda_H$ and $\Lambda_{\text{br}}$ can be achieved from a large ratio of the  periodicities $F/f$. 

The so-called clockwork mechanism~\cite{Choi:2014rja,Choi:2015fiu,Kaplan:2015fuy} gives a calculable example where $F/f\gg1\,$. This construction introduces $N+1$ spontaneously broken abelian symmetries at different sites of a moose diagram. The different sites are connected by $\epsilon$-suppressed operators, breaking explicitly $N$ of the abelian symmetries. The full clockwork potential is  

\begin{align}
&V_{\text{clock}}=\sum_{j=0}^{N}-m^2\vert\Phi_j\vert^2+g_{\text{clock}}^2\vert\Phi_j\vert^4+\Delta V_{\text{clock}}\,,\label{eq:clock_pot}\\
&\Delta V_{\text{clock}}=-\sum_{j=0}^{N-1}\left[\epsilon\Phi_{j}^\dagger\Phi_{j+1}^3+\text{h.c.}\right]\,,\label{eq:clockportal}
\end{align}
where the symmetry breaking scale is $f=m/g_{\text{clock}}\,$. By taking $\epsilon\ll g_{\text{clock}}^2 \sim 1\,$, the radial modes can be decoupled, and we can write an effective action for the $N+1$ angular modes $\pi_j$ by setting $\Phi_j=\frac{f}{\sqrt{2}} e^{i\pi_j/f}\,$. The  $\epsilon$-suppressed operators induce masses $m_j^2\approx \epsilon f^2$ for $N$ angular modes, stabilizing their VEVs at the origin. At the bottom of the spectrum, we are left with a single massless Nambu-Goldstone boson $\phi$, which non-linearly realizes the $U(1)_{\text{clock}}$ symmetry: $\pi_j\to \pi_j+\frac{1}{3^{j}} f\alpha\,$, $\phi\to \phi+ f\alpha\,$.

Since the overlap of $\phi$ with the site $j$ is suppressed by $\langle\pi_j |\phi\rangle\approx 1/3^{j}\,$, introducing an explicit breaking of $U(1)_{\text{clock}}$ at the site $j$ generates a potential for $\phi$ with periodicity of order $3^j f$. The desired hierarchy between the periodicities of the relaxion potential is then achieved by putting the backreaction sector at the $0$-\emph{th} site, and the rolling sector at the $N$-\emph{th} site. All in all, we get $F/f\approx 3^N\,$.
  
As first discussed in Ref.~\cite{Graham:2015cka}, by requiring a successful cosmological evolution for the relaxion, one gets a bound on the UV cutoff scale~\footnote{This bound characterizes relaxion scenarios where the Hubble friction during inflation is the main source of energy dissipation, and can be circumvented if other sources of energy dissipation, like for example particle production, become dominant (see Refs.~\cite{Hook:2016mqo,Tangarife:2017vnd,Tangarife:2017rgl}).} 
\begin{align}\label{eq:uppercut}
\!\!\!\!\!\Lambda_H&\lesssim \left(\frac{M_{\rm Pl}}{r_{\text{roll}}}\right)^{1/2}\cdot \left(\frac{\Lambda_{\text{br}}^4}{f}\right)^{1/6}\!\!\!\\
\!\!\!\!\!&\lesssim 10^{9}\text{ GeV}\cdot\!\left(\!\frac{1/4\pi}{r_{\text{roll}}}\!\right)^{\!\!1/2}\!\!\!\!\!\cdot\left(\frac{10^9\text{ GeV}}{f}\right)^{\!\!1/6}\!\!\!\!\!\cdot\left(\!\frac{\Lambda_{\text{br}}}{m_{h}}\!\right)^{\!\!2/3}\,,\!\!\!\notag
\end{align}
where to obtain the second inequality, we saturated both the upper bound on $\Lambda_{\text{br}}$ and the lower bound on $r_{\text{roll}}$. From Eq.~\eqref{eq:deriv}, we see that to get $\Lambda_H\sim 10^9\text{ GeV}\,$, one needs roughly $50$ clockwork sites.

Let us further comment on Eq.~\eqref{eq:order1}, which is at the origin of the relaxion CP problem. This is a generic prediction of the relaxion setup, which is intimately related to its dynamics. Indeed, having  $\vert\mu(\phi_0)\vert=m_h$ at the stopping point implies $\phi_0\approx F\,$. Therefore, a phase $\theta_N\equiv\frac{\phi_0}{F}\sim\mathcal{O}(1)$  is always induced by the rolling potential at the $N$-\emph{th} site of the clockwork. Generically, this gives also a phase $\theta_0\equiv\frac{\phi_0}{f}\sim\mathcal{O}(1)$ (modulo 2$\pi$) at the $0$-\emph{th} site. Therefore, \emph{if} the relaxion is identified with the QCD axion, and the backreaction potential is generated through a non-zero ABJ anomaly of $U(1)_{\text{clock}}$ with QCD, we have $\Lambda_{\text{br}}\sim(v\Lambda_{\textrm{QCD}}^3)^{1/4}\,$, and the $\mathcal{O}(1)$ phase $\theta_0$ induces a QCD $\theta$-angle of $\mathcal{O}(1)$, which is experimentally excluded. Notice that in the QCD case, $V_{\text{br}}$ depends linearly on the EW scale $v$ through the Gell-Mann-Oakes-Renner relation for the pion mass squared~\cite{GellMann:1968rz}, and the parametrization in Eq.~\eqref{eq:back} breaks down.

If one insists on generating $V_{\text{br}}$ via QCD, the relaxion cosmological dynamics should be modified in order to circumvent Eq.~\eqref{eq:order1}, and solve the strong CP problem. This is the approach followed in Ref.~\cite{Graham:2015cka}, and more recently in Refs.~\cite{Nelson:2017cfv,Jeong:2017gdy}. A trivial solution of the relaxion CP problem is instead to assume $U(1)_{\text{clock}}$ to be anomaly free with respect to QCD, and to generate the backreaction potential otherwise~\footnote{This includes also the setup presented in Ref.~\cite{Hook:2016mqo}, where the EW scale is selected by the particle production mechanism, and there is no need for a backreaction sector at all.}. However, this approach leaves generically the strong CP problem unaddressed, reducing the appeal of the original relaxion proposal.

In what follows, we show a third type of solution which assumes CP to be a symmetry of the UV theory, and $U(1)_{\text{clock}}$ to be anomaly free. The rolling potential is generated by a NB sector like the one presented in Refs.~\cite{Bento:1991ez,Dine:2015jga}, the phase $\theta_N$ is mapped into the CKM phase, and the strong CP problem is solved without modifying the standard relaxion dynamics. As a small drawback, the backreaction sector becomes less minimal - it requires extra states below $4\pi v$, and $r_{\text{br}}$ in Eq.~\eqref{eq:back} to be small enough to suppress the Higgs-independent wiggles (see Refs.~\cite{Graham:2015cka,Espinosa:2015eda,Gupta:2015uea,toappear} for different examples of working backreaction models).

\medskip

\section{The Nelson-Barr Relaxion}
We now present a simple implementation of the NB relaxion. The NB sector is borrowed from Ref.~\cite{Bento:1991ez}
\begin{equation}
\mathcal{L}_{\text{NB}}=\left[y^{\psi}_{i}\Phi_{N}+\!\tilde{y}^{\psi}_{i}\Phi_{N}^{*}\right]\psi u_{i}^{c}+\mu \psi \psi^c+\text{h.c.}\,,\label{eq:NB}
\end{equation}
where we use a basis in which all couplings are real. The SM up sector, $\mathcal{L}_{Y^u}=Y^u_{ij} HQ_i u^c_j+\textrm{h.c.}$, gets extended by two additional vector-like Weyl fermions, $\psi$ and $\psi^c$, in the fundamental and anti-fundamental of $SU(3)_C$, and with opposite hypercharges $\pm2/3$. The single complex scalar that breaks CP spontaneously in the NB model of Ref.~\cite{Bento:1991ez} is here identified with $\Phi_N$ at the $N$-\emph{th} site of the clockwork moose diagram. The $U(1)_N$  gets explicitly broken by the interactions of $\Phi_N$.  

The structure of the renormalizable couplings of $\Phi_N$ in Eq.~\eqref{eq:NB} is enforced by a $\mathbbm{Z}_2$ symmetry under which $\Phi_N$, $\psi$, and $\psi^c$ are charged. This discrete symmetry forbids operators like $\Phi_N\psi\psi^c$ and $H Q\psi^c$, and is spontaneously broken by the VEV of $\Phi_N$. Through the portal in Eq.~\eqref{eq:clockportal}, all the scalars VEV in the clockwork chain will break the $\mathbbm{Z}_2$ symmetry spontaneously.

In our minimal setup, the rolling potential for the relaxion $\phi$ is generated from Eq.~\eqref{eq:NB}. Matching to the potential in Eq.~\eqref{mu1}, we get 
\begin{align}
\Lambda_H  &\sim\frac{\sqrt{y^{\psi}_{i} \tilde{y}^{\psi}_{j}(Y^{u\dagger}Y^u)_{ij}}}{4 \pi} f\,,\label{eq:matching1}\\
r_{\text{roll}} &\sim \frac{4 \pi g_{\text{clock}} \sqrt{y^{\psi}_{k} \tilde{y}^{\psi}_{k}}}{y^{\psi}_{i} \tilde{y}^{\psi}_{j}(Y^{u\dagger}Y^u)_{ij}}\,,\label{eq:matching2}
\end{align}
where $g_{\text{clock}}$ is the clockwork coupling of Eq.~\eqref{eq:clock_pot}. In the presence of the backreaction potential, the relaxion stops where $\theta_N\sim{\cal O}(1)\,$, as dictated by Eq.~\eqref{eq:order1}. Setting $\Phi_N$ to its VEV, we can define $B_i=\frac{f}{\sqrt{2}}\left(y^{\psi}_{i}e^{i\theta_N}+\tilde{y}^{\psi}_{i}e^{-i\theta_N}\right)\,$. The $4\times4$ mass matrix of the up quarks at tree-level is
\begin{equation}\label{eq:mass_D_tree_level}
M^{u}=\left(\begin{array}{cc}
	(\mu)_{1\times1} & \left(B\right)_{1\times3}\\
	(0)_{3\times1} & \left(vY^u\right)_{3\times3}
\end{array}\right)\,,
\end{equation}
so that, even though the above mass matrix is complex, we find $\bar{\theta}_{\textrm{QCD}}={\rm Arg}({\rm det}(M^{d}))+{\rm Arg}(\mu\cdot{\rm det}(vY^u))=0\,$.
Integrating out the heavy quarks for $\mu^{2}+B_{i}B_{i}^{*}\gg v^2$ (where here and below $i,j,k,\ell=1..3$ are flavor indices), we find the effective $3\times3$ mass squared of the SM up quark sector:
\begin{equation}\label{iout}
\left[M^{u}_{\textrm{eff}}M^{u\dagger}_{\textrm{eff}}\right]_{ij}\sim v^2 Y^{u}_{ik} Y^{u *}_{jk}-\frac{v^2Y^{u}_{ik}B_{k}^{*}B_{\ell}Y^{u *}_{j\ell}}{\mu^{2}+|B|^2}\,.
\end{equation}
The phase in the unitary matrix $V_u^L$, which diagonalizes the matrix in Eq.~\eqref{iout}, would lead to a phase, $\delta_{\rm CKM}$, in the CKM matrix, $V_{\rm CKM}=V_u^{L\dagger} V_d^L\,$. It is straightforward to show that in the limit $|\vec y^{\psi} \times \vec {\tilde{y}}^{\psi}| / |\vec y^{\psi} + \vec {\tilde{y}}^{\psi}|^2\ll1\,$, the CKM phase vanishes.
To ensure $\delta_{\rm CKM}\sim\mathcal{O}(1)\,$, we assume for simplicity $y^{\psi}_{i}\sim\tilde{y}^{\psi}_{i}\sim y_\psi$ for all $i$, and $v<\mu \lesssim |B_i|\sim |y_\psi| f\,$. The latter requirement can be, in principle, relaxed, since $\mu$ is a technically natural parameter which can be set within a large range of values between $v$ and $y_\psi f$. In the limit where $\mu\ll |B_i|\,$, one of the eigenvalues of the up quark mass matrix squared in Eq.~\eqref{iout} is suppressed by $\mu^2/\vert B_i\vert^2\,$, and can possibly address part of the flavor puzzle (say accounting for the smallness of the up quark mass). We leave a study of this possibility for future work, and take $\mu\sim |B_i|\,$. 

A careful analysis of the radiative corrections to our construction is left to the supplementary material. Our main results are in agreement with Ref.~\cite{Dine:2015jga}, since the leading order deviation from the NB construction are captured by the same set of higher dimensional operators. The strong experimental upper bound on $\bar{\theta}_{\textrm{QCD}}$ translates into an upper bound on $y_i^{\psi}$ and $\tilde{y}_i^{\psi}$:
\begin{equation}\label{eq:bound}
\Delta\bar{\theta}_{\textrm{QCD}}^{UV}\lesssim10^{-10}\quad \Rightarrow\quad \vert y_i^\psi\vert\sim\vert\tilde{y}_i^\psi\vert\lesssim 10^{-2},\,10^{-4}\,.
\end{equation}
We quote above two bounds, depending on the flavor structure assumed - the weaker [stronger] is related to models with minimal flavor violation~(MFV) [quasi-diagonal] structure (see auxiliary material for details). As we will see in the next section, this bound will have important consequences on the parameter space of the NB relaxion.

In our construction, the cut off of the Higgs sector is much smaller than the one of the clockwork theory, $\Lambda_H\ll \Lambda_f\,$. Our analysis of the radiative stability of the NB construction does not include higher dimensional operators arising at the IR threshold $\Lambda_H$. These contributions are model dependent, and could, in principle, lead to bigger contributions to $\Delta\bar{\theta}_{\textrm{QCD}}$ than the ones considered here. A proper estimation would require a concrete UV completion of the NB relaxion, embedding the Higgs IR threshold, $\Lambda_H$, in a full model of SUSY or Compositeness. These UV completions are challenging, and beyond the scope of this work (see Refs.~\cite{Batell:2015fma,Batell:2017kho,Evans:2017bjs} for attempts of relaxion UV completions, and also Refs.~\cite{Dine:1993qm,Dine:2015jga,Vecchi:2014hpa} for SUSY or Composite UV complete NB models).

Within our simple model, the radiative contributions to $\Delta\bar{\theta}_{\textrm{QCD}}$ of the new states at $\Lambda_H$ can be subleading with respect to the ones considered here if the interactions of these states with the clockwork chain are suppressed enough. A better solution would be instead to modify the NB construction by softening the breaking of the $U(1)_{\text{clock}}$, such that the cutoff scale of the rolling potential is sequestered with respect to $f$ (this can be achieved in models like the ones in Ref.~\cite{Gupta:2015uea}). With such a construction, one could possibly achieve $\Lambda_H\sim\Lambda_f\,$, suppressing the higher dimensional operators at $\Lambda_H$ as much as the ones we considered here. We leave this matter for future investigations, and focus here on the simplest possible model, where we introduce a brute force explicit breaking of the $U(1)_N$, like in Eq.~\eqref{eq:NB}, which leads to the ``quadratically-divergent" rolling potential of Eq.~\eqref{eq:matching2}.

\section{Phenomenology}

\begin{figure*}[!t]
	\includegraphics[width=0.5\textwidth]{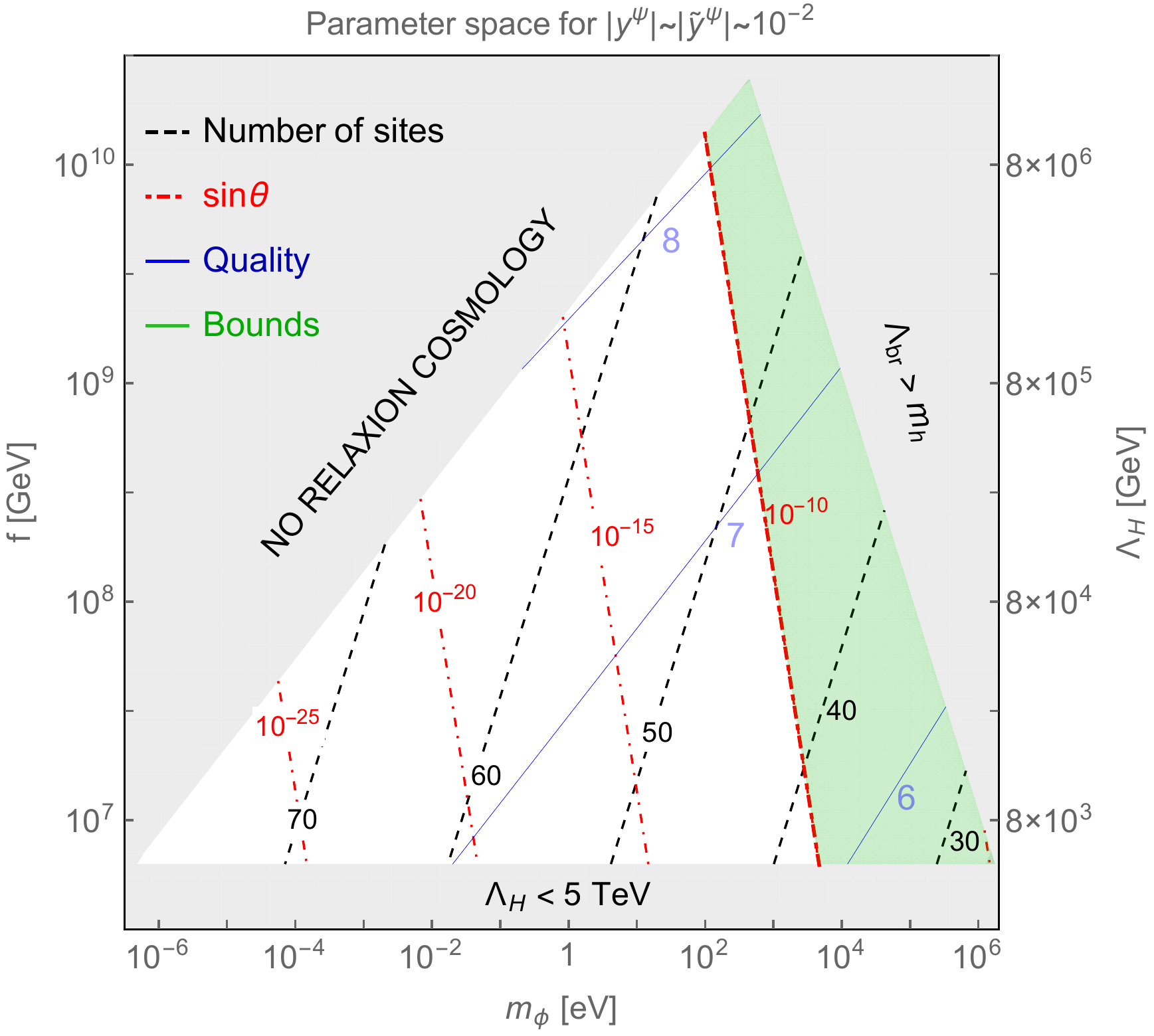}\hfill
	\caption{The white region shows the allowed parameter space of the Nelson-Barr relaxion. The black dashed contours give the number of clockwork sites, the red dot-dashed contours correspond to the relaxion-Higgs mixing, and the thin blue contours indicate the minimal dimension of the Planck-suppressed operators that should be forbidden. The colored-shaded region combines current constraints from astrophysics~\cite{Raffelt:1987yt,Raffelt:2012sp,Hardy:2016kme,Chang:2018rso}, Extragalactic Background Light~\cite{Flacke:2016szy}, and flavor-violating Kaon decays~\cite{Adler:2008zza} (see Appendix for further details).}
	\label{fig:summary}
\end{figure*}

We now assess the parameter space of the NB relaxion, and study its phenomenology.  In our minimal setup, the rolling potential is generated from the NB sector at the $N$-\emph{th} site. As explained above, this implies that the cutoff $\Lambda_H$ is related to the scale $f$. Taking Eq.~\eqref{eq:matching1} at face value, given a lower bound on the new physics scale $\Lambda_H\gtrsim 5\text{ TeV}$, and the upper bound on the explicit breaking of the $U(1)_N$ in Eq.~\eqref{eq:bound}, we get 
\begin{equation} 
f^{\text{min}}=\left(10^{7},\,10^{9} \right)\text{ GeV}\cdot\left(\frac{\Lambda_H}{5\,\text{TeV}}\right)\cdot\frac{\left(10^{-2}, \,10^{-4} \right)}{\sqrt{y^{\psi}\tilde{y}^{\psi}}} \label{eq:cutoffbound2}\,,
\end{equation}
where the weaker (stronger) lower bound is related to models with MFV (quasi-diagonal) flavor structure. From now on, we are going to focus on the MFV parameter space, which is presented in Fig.~\ref{fig:summary}, and leave the discussion on the quasi-diagonal case for the additional material.

The mass of the relaxion is controlled by the backreaction scale $m_\phi\approx \Lambda_{\text{br}}^2/f\,$, so that the upper bound in Eq.~\eqref{eq:boundback} gives an upper bound on the relaxion mass, which reads
$m_\phi^{\text{max}}=20\text{ MeV}\cdot\frac{5\text{ TeV}}{\Lambda_H}\cdot\frac{\sqrt{y^{\psi}\tilde{y}^{\psi}}}{10^{-2}}\,$.
The constraint in Eq.~\eqref{eq:uppercut}, coming from a successful relaxion cosmology, gives instead a lower bound on the relaxion mass
$m_\phi^{\text{min}}=2\cdot10^{-6}\text{ eV}\cdot\left(\frac{\Lambda_H}{5\text{ TeV}}\right)^{5/2}\cdot\frac{10^{-2}}{\sqrt{y^{\psi}\tilde{y}^{\psi}}}\,$.
The three constraints above explain the boundaries of the allowed triangle in Fig.~\ref{fig:summary}. The maximal allowed decay constant is achieved for $m_\phi^{\text{min}}=m_\phi^{\text{max}}\,$, and reads
$f^{\text{max}}=3\cdot 10^{10}\text{ GeV}\cdot\left(\frac{10^{-2}}{\sqrt{y^{\psi}\tilde{y}^{\psi}}}\right)^{3/7}\,$. The available parameter space  shrinks for smaller $\sqrt{y^{\psi}\tilde{y}^{\psi}}$ until $f^{\text{max}}\approx f^{\text{min}}$, and there is no more parameter space left. This happens for $\sqrt{y^{\psi}\tilde{y}^{\psi}}\sim 10^{-9}$, which can be taken as the \emph{minimal} amount of $U(1)_N$ breaking in our setup.

Through the backreaction sector, a relaxion-Higgs mixing is generated. The mixing angle, $\sin\theta\approx7\cdot10^{-7}\frac{f}{\text{ GeV}}\cdot\frac{m_\phi^2}{\text{ GeV}^2}\,$, is plotted as red contours in Fig.~\ref{fig:summary}. Within our range of parameters, the relaxion is always long-lived on detector scale (being lighter than 20 MeV, its dominant decay modes are into electron and photon pairs, which are heavily suppressed). Following the analysis in Ref.~\cite{Flacke:2016szy}, we plot in Fig.~\ref{fig:summary} an exclusion band which comes from astrophysical probes~\cite{Raffelt:1987yt,Raffelt:2012sp,Hardy:2016kme,Chang:2018rso}, and flavor experiments~\cite{Adler:2008zza}. More details are given in the supplementary material.

In principle, another interesting feature of our setup is the changing of $\delta_{\text{CKM}}\approx\langle\phi\rangle/F$ during the relaxion rolling. However, this would hardly lead to observable effects, unless the flavor suppression of the SM CP violation is somehow reduced~\cite{Perez:2005yx}.

\section{Quality problems}
The spontaneous breaking of the global $U(1)_{\text{clock}}$ at a high scale introduces a ``relaxion quality problem'', which shares some similarities with the ``axion quality problem'' discussed in Ref.~\cite{Kamionkowski:1992mf}. The basic issue is that gravity is not expected to respect any global symmetry, as suggested by many theoretical arguments~\cite{Abbott:1989jw,Coleman:1989zu,Kallosh:1995hi,Banks:2010zn}. As a consequence, gravity-induced higher dimensional operators would generically break the $U(1)_{\text{clock}}$, generating a potential for the  relaxion. The most dangerous gravity contributions to the potential are the ones controlled by operators like $\frac{\Phi_0^{4+\Delta}}{M_{\rm Pl}^{\Delta}}$, with $\Delta\geq1$ (see Ref.~\cite{Higaki:2016yqk} for an early discussion on the quality problem of the clockwork construction). These operators generate $\Delta V_{\text{grav}}\sim \frac{f^{4+\Delta}}{M_{\rm Pl}^\Delta}\cos\frac{\Delta \phi}{f}\,$, which, being independent of the Higgs VEV, should be smaller than the backreaction potential in Eq.~\eqref{eq:back} in order for the relaxion mechanism to work~\footnote{The breaking of global symmetries arises already at the perturbative level in calculable string theory setups (see Ref.~\cite{Banks:1988yz}). For this reason, we parametrize the breaking of the global symmetry by local Planck-suppressed operators. We refer to Ref.~\cite{Alonso:2017avz} for a discussion on subleading non-perturbative contributions from gravitational instantons.}. By imposing  $\Delta V_{\text{grav}}\lesssim V_{\text{br}}$ together with Eq.~\eqref{eq:boundback} and Eq.~\eqref{eq:cutoffbound2}, we get
$\Delta\!\gtrsim\!\left[\frac{4\log\frac{\Lambda_{\text{br}}}{f}}{\log\frac{f}{M_{\rm Pl}}}\right]\!\approx 2\,$,
which implies that gravity-induced operators up to dimension 6 should be forbidden. This is a generic problem of every relaxion model with high decay constant $f$. Since imposing gauged $\mathbbm{Z}_N$ symmetries on the gravity theory seems challenging in the clockwork setup, the only way of addressing this problem would be to build a UV completion where the $U(1)_{\text{clock}}$ is an accidental symmetry emerging in the infrared as a result of the UV gauge symmetries. This last possibility has been  explored in the context of clockwork constructions in Refs.~\cite{Kaplan:2015fuy,Coy:2017yex}.

The NB sector of our construction can also be affected by gravity-induced higher dimensional operators. These operators evade our power counting because they are not controlled by powers of $y_\psi$, and they break MFV, leading to dangerous contributions to $\bar{\theta}_{\textrm{QCD}}$. If dimension 5 operators are forbidden, the effects of dimension 6 operators are already small enough for $f\sim 10^{9}\text{ GeV}$ to guarantee a successful NB mechanism. Of course, raising $f$ would make the screening of Planck-suppressed operators more challenging, making it necessary to embed the NB construction in a gauge theory where the CP violating phase arises from a condensate (see for example Refs.~\cite{Dashen:1970et,Vecchi:2014hpa,Gaiotto:2017tne}).

\section{Conclusions}

The main lesson of this work is that combining the relaxion mechanism with the Nelson-Barr~(NB) construction leads to two positive outcomes: (i)~the ``relaxion CP problem'', induced by the $\mathcal{O}(1)$ CP phase of the relaxion vacuum expectation value, is solved; (ii)~the relaxion CP phase becomes the origin of the Cabibbo-Kobayashi-Maskawa~(CKM) phase. Our model serves as an existence proof of the NB relaxion setup, focusing on the simplest possible implementation, which captures some generic features of the construction. We showed how the NB sector provides the relaxion ``rolling'' potential, connecting the relaxion decay constant with the cutoff scale, where new physics stabilizing the Higgs mass is expected.


Within our setup, the cosmological evolution of the relaxion is kept as minimal as in the original proposal~\cite{Graham:2015cka}. This should be contrasted with other proposals to solve the relaxion CP problem, which either require the relaxion potential to be modified after inflation~\cite{Graham:2015cka}, or the classical evolution of the relaxion to be overcome by its large quantum fluctuations~\cite{Nelson:2017cfv}.

A feeble coupling between the NB sector and the relaxion seems necessary in order to guarantee the success of the construction. This makes the relaxion detection challenging, even though signatures are expected for maximal backreaction scale in astrophysical and cosmological observables, and Kaon experiments.

Breaking the $U(1)_{\text{clock}}$ at a high scale introduces the theoretical difficulty of protecting the entire construction from Planck-suppressed operators. It would be interesting to see if any of these problems can be ameliorated in more elaborate versions of the NB relaxion, which include Supersymmetry or Compositeness to stabilize the UV cutoff (see Refs.~\cite{Dine:2015jga,Vecchi:2014hpa} for a discussion on the NB mechanism in these frameworks).

The whole NB construction is deeply connected to the flavor texture of the quark mass matrix. It is then tempting to think about embedding the NB relaxion in a full model of flavor, where the $U(1)_{\text{clock}}$ plays the role of a horizontal flavor symmetry, and the mass hierarchies and structure of the CKM are explained \emph{\`a la} Froggatt-Nielsen~\cite{Froggatt:1978nt}. We will show how this can be done in a companion paper~\cite{toappear}.

\medskip

\subsection*{Acknowledgements}
We thank David Kaplan, Zohar Komargodski, Yossi Nir, Lorenzo Ubaldi, Tomer Volansky, and Jure Zupan for useful discussions. We also thank Nathaniel Craig and Claudia Frugiuele for comments on the draft, and Alfredo Urbano and Michael Dine for fruitful discussions on the quality problem.  The work of GP is supported by grants from the BSF, ERC, ISF, Minerva, and the Weizmann-UK Making Connections Program.  \medskip

\medskip

\appendix

\section{Radiative Stability of Nelson-Barr}
We comment here on the radiative stability of the NB construction we presented. Dangerous radiative corrections can be parametrized in a shift of the mass matrix of the up sector, that we schematically write as $M^u\to M^u+\Delta M^u\,$, where $\Delta M^u \ll M^u$ is
\begin{equation}\label{eq:mass_matrix}
\Delta M^u\equiv\!\!\left(\!\begin{array}{cc}
\left(\delta\mu\right)_{1\times1} & \left(\delta B\right)_{1\times3}\\
\left( v\delta Y^{\psi^{c}}\right)_{3\times 1}  & \left(v\delta Y^u\right)_{3\times3}
\end{array}\!\right)\,,
\end{equation}
and its contribution to $\bar{\theta}_\textrm{QCD}$ reads
\begin{equation}\label{eq:Tr_identity}
\Delta\bar{\theta}_{\textrm{QCD}}\approx\text{Im}\left[\text{Tr}\!\left(\!\left(M^u\right)^{\!-1}\!\!\Delta M^u\!\right)\right]\,.
\end{equation}
There are three different types of contributions 
\begin{align}
\delta\bar{\theta}_{1} & = \text{Im}\left[\mu^{-1}\delta\mu\right]\,,\label{eq:delta theta 1}\\
\delta\bar{\theta}_{2} & = -\text{Im}\left\{\text{Tr}\left[\mu^{-1}B\left(Y^{u}\right)^{-1}\delta Y^{\psi^{c}}\right]\right\}\,,\label{eq:delta theta 2}\\
\delta\bar{\theta}_{3} & = \text{Im}\left\{\text{Tr}\left[\left(Y^u \right)^{-1}\delta Y^{u}\right]\right\}\,.\label{eq:delta theta 3}
\end{align}
By promoting the couplings $y^{\psi}_{i}$, $\tilde{y}^{\psi}_j$, and $\mu$ to spurions, one can see from Table~\ref{table:spurion} that there is a (set of) charge assignment(s) of three $U(1)$'s, acting non-trivially on $\Phi_N$ and the vector-like fermions. By assuming that the full theory respects the same selection rules, we can parametrically estimate by  spurion counting the leading order contributions to $\bar{\theta}_{\textrm{QCD}}$. 

\begin{table}[tb]
	\begin{tabular}{c c|| c c c}
		
		& $\quad\mathbbm{Z}_2\quad$ &$U\left(1\right)_{N}$& $U\left(1\right)_{\psi}$ & $U\left(1\right)_{\mu}$ \\ [0.5ex]
		\hline
		$\Phi_{N}$& -& -$1$ & 0 & 0 \\
		$\psi$ & -& $0$ & 1 & 0 \\
		$\psi^{c}$ &-& $0$ & -1 & 1 \\
		\hline\hline
		$y_{i}^{\psi}$ &+& $1$ & -1 & 0 \\
		$\tilde{y}_{i}^{\psi}$&+& -$1$ & -1 & 0 \\
		$\mu$&+ & 0 & 0 & -1 \\ [1ex]
	\end{tabular}
	\caption{Charges of the spontaneously broken $\mathbbm{Z}_2$. Spurion charge assignment for the three broken $U(1)$ symmetries controlling the radiative corrections}
	\label{table:spurion}
\end{table}

We first estimate the contributions from integrating out the clockwork scalar chain. Inspecting Table~\ref{table:spurion}, we conclude that the ${\cal O}(y^\psi \tilde y^\psi )$ contributions to Eq.~\eqref{eq:delta theta 1} should be proportional to $\text{Im}\left[y^{\psi}_i\tilde{y}^{\psi*}_i\!\langle \Phi_{\!N}\rangle^{\!2}\!\!+\!y^{\psi*}_i\tilde{y}^{\psi}_i\!\langle \Phi_{\!N}^*\rangle^{\!2}\right]\!\!=\!0\,$, where we have used the fact that our Lagrangian is invariant under the interchange $\left(y^{\psi}_{i}, \Phi_{N}\right) \leftrightarrow \left(\tilde{y}^{\psi}_{i}, \Phi_{N}^{*}\right)\,$. The contributions to Eq.~\eqref{eq:delta theta 2} and Eq.~\eqref{eq:delta theta 3} similarly vanish at the same order, and no misalignment can be produced to get a non-zero phase.

\begin{figure*}[!t]
	\includegraphics[width=0.5\textwidth]{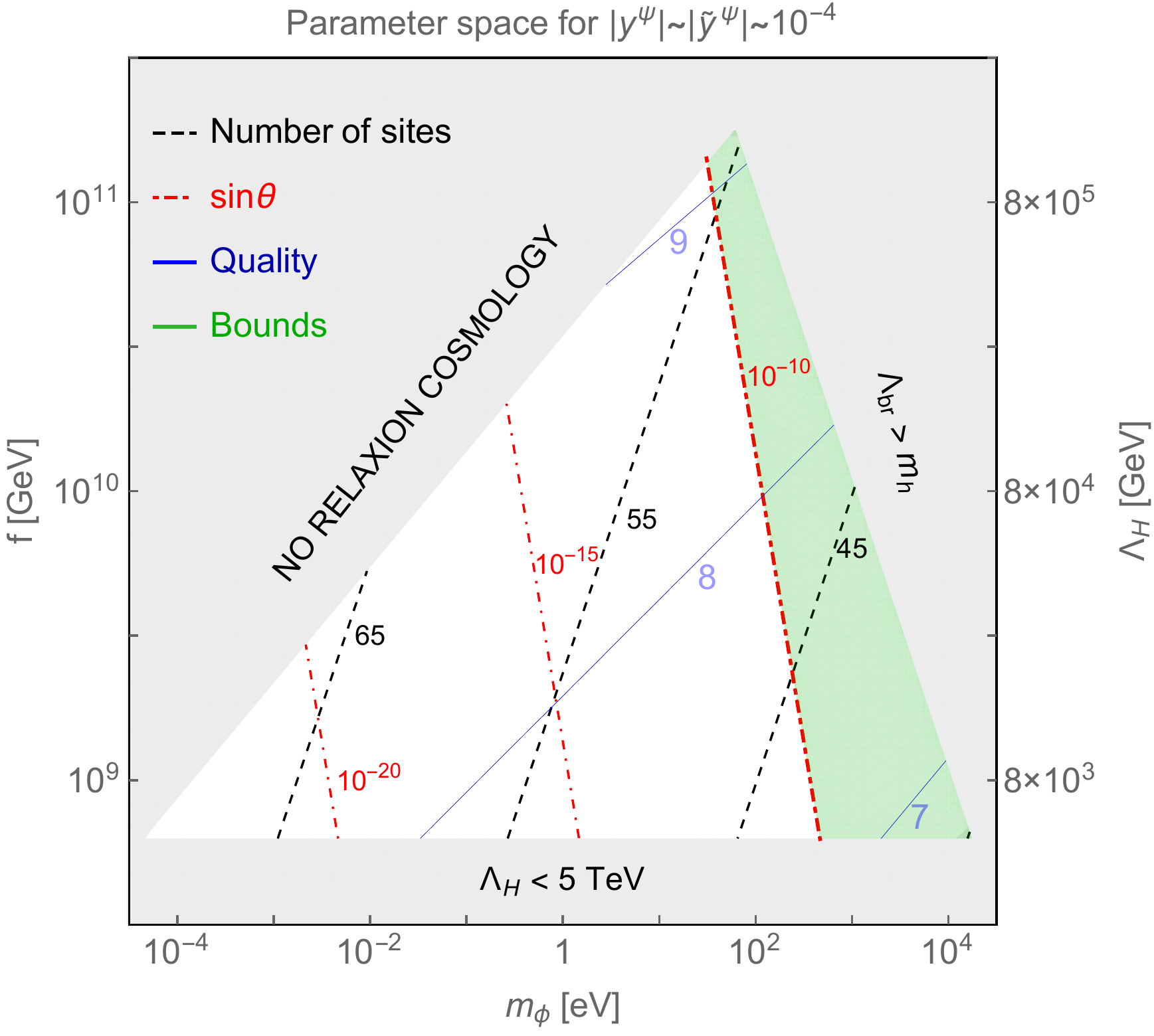}\hfill
	\caption{The white region shows the allowed parameter space of the Nelson-Barr~(NB) relaxion, in the ($m_{\phi}\, ,\, f$) plane. The black dashed contours indicate the number of clockwork sites required to achieve the desired hierarchy. The red dot-dashed contours indicate the mixing between the relaxion and the SM Higgs. The thin blue contours indicate the minimal dimension of the Planck-suppressed operators that should be forbidden.}
	\label{fig:summary2}
\end{figure*}

Our spurion analysis is in agreement with the explicit computation of Refs.~\cite{Bento:1991ez,Dine:2015jga}, which shows that the first dangerous loop corrections in this model are controlled by Higgs-portal operators like $H^{\dagger}H \Phi_N^2$. These are generated at 1-loop in our setup, together with the rolling potential for the relaxion, only after integrating out the heavy quarks. As a consequence, the leading contributions to Eqs.~(\ref{eq:delta theta 1}-\ref{eq:delta theta 3}) arise at 2-loops, and their sum scales as
\begin{equation}
\Delta\bar{\theta}_\textrm{QCD}^{\text{clock}}\sim\frac{(y^{\psi}\tilde{y}^{\psi})^2}{256 \pi^4}\!\frac{\left|\langle \Phi_N\rangle\right|^2}{m^2}\sin\left(2\theta_N\right)\,,
\end{equation}
where $m$ is the mass of the radial mode in the clockwork construction. The upper bound on $\bar{\theta}_{\text{QCD}}$  translates into an upper bound on the explicit breaking of the $U(1)_N$:
\begin{equation}\label{eq:boundless}
\Delta\bar{\theta}_{\textrm{QCD}}^{\text{clock}}\lesssim10^{-10}\quad \Rightarrow\quad \vert y^\psi\vert\sim\vert\tilde{y}^\psi\vert\lesssim 10^{-2}\,.
\end{equation}

We now consider the effects of higher dimensional operators. The leading order higher dimensional operators generated at the cutoff scale of the clockwork theory, $\Lambda_f\sim4\pi f\,$, are 
\begin{align}
\mathcal{L}_{\text{UV}}&\supset\frac{\tilde{y}^{\psi *}_j}{16\pi^2}\left[y^{\psi}_i\alpha_{ij}\mu\frac{\Phi_N^2}{f^2}\psi\psi^c+\beta_{kj}\mu\frac{\Phi_N}{f^2}HQ_{k}\psi^c\right.\notag\\
&\left.\qquad\qquad+y^{\psi}_i\gamma_{k\ell ij}\frac{\Phi_N^2}{f^2}HQ_{k}u_{\ell}^c\right]+\text{h.c.}\,,\label{eq:highedim}
\end{align}
and other operators obtained by interchanging $\left(y^{\psi}_{i}, \Phi_{N}\right) \leftrightarrow \left(\tilde{y}^{\psi}_{i}, \Phi_{N}^{*}\right)\,$. The first operator induces a contribution to $\bar{\theta}_{\text{QCD}}$  which can be directly estimated through Eq.~\eqref{eq:delta theta 1}
\begin{equation}
\Delta\bar{\theta}_\textrm{QCD}^{\text{UV}}\!\sim\!\frac{\alpha_{ij}}{16 \pi^2}\!\left(y^{\psi}_i\tilde{y}^{\psi}_j-y^{\psi}_j\tilde{y}^{\psi}_i\right)\!\frac{\left|\langle \Phi_N\rangle\right|^2}{f^2}\sin\left(2\theta_N\right)\,,\label{eq:leading}
\end{equation}
and we again assumed the UV Lagrangian to be invariant under the interchange $\left(y^{\psi}_{i}, \Phi_{N}\right) \leftrightarrow \left(\tilde{y}^{\psi}_{i}, \Phi_{N}^{*}\right)\,$. In realizations of Minimal Flavor Violation~(MFV)~\cite{DAmbrosio:2002vsn}, the coefficient $\alpha_{ij}$ is either proportional to the identity or to powers of $(Y^{u\dagger} Y^{u})_{ij}$. In either cases, the contribution to $\Delta\bar{\theta}_\textrm{QCD}^{\text{UV}}$ in Eq.~\eqref{eq:leading} vanishes. Higher power in $(Y^{d\dagger} Y^{d})_{ij}$ might lead to violations of this scaling~\cite{Kagan:2009bn,Barbieri:2011ci}, but will be highly suppressed. An analogous reasoning applies to the operators controlled by $\beta$ and $\gamma$, so that the contribution to $\Delta\bar{\theta}_\textrm{QCD}^{\text{UV}}$ at leading order in $y_{[i}^\psi\tilde{y}^\psi_{j]}$ identically vanishes. The first non-zero contribution is then of order $\mathcal{O}(y_{\psi}^4)$, and hence suppressed compared to the generic case as in Eq.~\eqref{eq:boundless}.

Notice that the other two operators in Eq.~\eqref{eq:highedim} give contributions of the same order of the ones from Eq.~\eqref{eq:leading} only because the factors $\left(Y^u\right)^{-1}\beta$ and $\left(Y^u\right)^{-1}\gamma$, in Eq.~\eqref{eq:delta theta 2} and Eq.~\eqref{eq:delta theta 3}, do not lead to any enhancement in MFV scenarios. More generally, this enhancement is not present in a large class of flavor models where the couplings of the first two generations are suppressed, resulting in quasi-diagonal textures. This is the case in various $U(2)$ models, in $U(1)$ horizontal models, and in models where the flavor puzzle is addressed through hierarchies in the anomalous dimensions (for a discussion, see {\it e.g.} Refs.~\cite{KerenZur:2012fr,Barbieri:2011ci,Sala:2012ib,Sala:2012xy,Panico:2015jxa,Isidori:2010kg,Gedalia:2010rj} and Refs. therein). In this class of models, the contributions of order $\mathcal{O}(y_\psi^2)$, such as the ones in Eq.~\eqref{eq:leading}, do not vanish, resulting in the stronger bound $y_\psi \lesssim 10^{-4}\,$.

\section{More details on the phenomenology}
The flavor-violating invisible decay of the Kaon is bounded at the  90\% CL to be $\text{BR}(K^{+}\to\pi+\phi)< 7.3\cdot10^{-11}$ by combined data from E787 and E949 experiments \cite{Adler:2008zza}. The astrophysical constraints arise from energy loss arguments for the SN1987a supernova. As first proposed in Ref.~\cite{Raffelt:1987yt}, we require the cooling rate into relaxion to be less than $6\cdot10^{55}\text{ GeV/s}$. We neglect possible uncertainties coming from the modeling of the neutrino emission from the collapse (see Refs.~\cite{Kushnir:2014oca,Kushnir:2015mca,Blum:2016afe}).  This bound suffers also from large systematical uncertainties due to the poorly known parameters of the supernova, like its temperature, its core radius, and the neutron density (we fix them, following Ref.~\cite{Ishizuka:1989ts}, to $T=60\text{ MeV}\,$, $R_{\text{core}}=10\text{ km}\,$, and $\rho_n=3\cdot 10^{14}\text{ g}/\text{cm}^3$). A more recent analysis \cite{Chang:2018rso} addresses these uncertainties, strengthening the robustness of this bound.

Further bounds can be derived from cooling rate of the Sun, Horizontal Branch stars, and Red Giants~\cite{PhysRevD.86.015001,Hardy:2016kme}. The late decays of the relaxion can also affect the diffuse Extragalactic Background Light, first computed in Ref.~\cite{Flacke:2016szy}. Present fifth force experiments~\cite{PhysRevLett.44.1645,PhysRevD.32.3084,Chiaverini:2002cb,Smullin:2005iv,Kapner:2006si} do not have the sensitivity of probing a sub-eV mediation with mixing with the SM Higgs smaller than $10^{-15}$.

We finally present, for completeness, the allowed parameter space for the NB relaxion when the Wilson coefficients of the higher dimensional operators in Eq.~\eqref{eq:highedim} do not satisfy MFV, but have quasi-diagonal texture. The upper bound on the spurions $y_i^{\psi}$ and $\tilde{y}_j^{\psi}$ is two orders of magnitude stronger in this case, resulting in a higher decay constant, $10^{9}\textrm{ GeV}\cdot\frac{\Lambda_H}{5\textrm{ TeV}}\cdot\frac{10^{-4}}{\sqrt{y^{\psi}\tilde{y}^{\psi}}}\lesssim f\lesssim 2\cdot 10^{11}\text{ GeV}\cdot\left(\frac{10^{-4}}{\sqrt{y^{\psi}\tilde{y}^{\psi}}}\right)^{3/7}\,$. The mass range of the NB relaxion is then $2\cdot10^{-4}\text{ eV}\cdot\left(\frac{\Lambda_H}{5\text{ TeV}}\right)^{5/2}\cdot\frac{10^{-4}}{\sqrt{y^{\psi}\tilde{y}^{\psi}}}\lesssim m_{\phi}\lesssim0.2\text{ MeV}\cdot\frac{5\text{ TeV}}{\Lambda_H}\cdot\frac{\sqrt{y^{\psi}\tilde{y}^{\psi}}}{10^{-4}}\,$. The available parameter space is presented in Fig.~\ref{fig:summary2}.

\medskip
\bibliography{hierarchion}

\end{document}